\RequirePackage{fix-cm}
\documentclass[smallextended]{svjour3}       
\smartqed  
\usepackage{graphicx}
\usepackage{amsmath}

\usepackage{enumerate}

\usepackage{multirow}

\usepackage{amsfonts,amssymb}

\usepackage{cite}
\begin{document}

\title{Novel Reconciliation Protocol Based on Spinal Code for Continuous-variable Quantum Key Distribution 
}


\author{ 	Xuan Wen \and Qiong Li \and Haokun Mao \and Yi Luo \and Bingze Yan 
}

\institute{	
	Qiong Li\\
	\email{qiongli@hit.edu.cn} \\
	Xuan Wen \and Qiong Li \and Haokun Mao \and Yi Luo \and Bingze Yan
	\at   Department of Computer Science and Technology, Harbin Institute of Technology, Harbin, China \\
}

\date{Received: date / Accepted: date}

\maketitle

\begin{abstract}
Reconciliation is a crucial procedure in post-processing of continuous variable quantum key distribution (CV-QKD) system, which is used to make two distant legitimate parties share identical corrected keys. The adaptive reconciliation is necessary and important for practical systems to cope with the variable channel. Many researchers adopt the punctured LDPC codes to implement adaptive reconciliation. In this paper, a novel rateless reconciliation protocol based on spinal code is proposed, which can achieve a high-efficiency and adaptive reconciliation in a larger range of SNRs. Due to the short codes length and simple structure, our protocol is easy to implement without the complex codes designs of fixed rate codes, e.g., LDPC codes. Meanwhile, the structure of our protocol is highly parallel, which is suitable for hardware implementation, thus it also has the potential of high-speed hardware implementation. Besides, the security of proposed protocol is proved in theory. Experiment results show that the reconciliation efficiency maintains around $95\%$ for ranging SNRs in a larger range (0,0.5), even exceeds $96.5\%$ at extremely low SNR ($\le 0.03$) by using this novel scheme. The proposed protocol makes the long-distance CV-QKD systems much easier and stable to perform a high-performance and adaptive reconciliation.

\end{abstract}

\section{Introduction}
\label{intro}
 Quantum key distribution (QKD) is the art of distributing provably unconditional security keys between two remote legitimate parties Alice and Bob by encoding information on photons, even if in the presence of an eavesdropper (Eve). The QKD protocols mainly contains discrete-variable (DV) protocols \cite{1,2,3,4,5,6,7} and continuous-variable (CV) protocols \cite{8,9,10,11,12,13}. DV-QKD protocols encode key information on discrete variables such as the phase or the polarization of single photons. CV-QKD protocols encodes key information on continuous variables such as the quadratures of coherent states. The limitation and technological challenges of DV-QKD mainly lie in the speed and efficiency of photon detectors in the single-photon regime. Compared to DV-QKD protocols, CV-QKD modulate and detect the coherent states efficiently by using standard telecommunication technologies, which allows one to eliminate these constraints of DV-QKD.

Generally, a CV-QKD system is carried out in two consecutive phases, namely, the quantum key establishment phase and the classical post-processing phase. In the former phase of a practical Gaussian-modulated coherent state CV-QKD system \cite{8}, Alice prepares coherent state using two random values \(X_\mathrm{A}\), \(P_\mathrm{A}\) with a Gaussian distribution \(N(0, V_A)\) and sends it to Bob via the quantum channel. Bob randomly chooses to measure one of the two quadratures of the received coherent state and then informs Alice of his choice of quadrature. Due to the physical noises or the presence of Eve \cite{14}, Alice and Bob share weakly correlated and insecure continuous-variable raw data. Thus, the latter phase called post-processing must be performed to extract identical secret keys from the correlated raw data via a classical authenticated channel. The post-processing contains four main stages: sifting, parameter estimation \cite{15,16}, reconciliation \cite{17,18,19,20} and privacy amplification \cite{21,22}. The reconciliation stage is aim to correct errors of the correlated raw data, which plays a crucial role in practical CV-QKD systems since its performance affects both the final secret key rate and transmission distance\cite{23,24,25}. Implementing high-performance reconciliation is the main bottleneck of post-processing for long-distance CV-QKD systems \cite{26}.

Up to now, a series of reconciliation schemes have been explored for CV-QKD system. The earliest scheme called sign reconciliation uses the sign of the correlated continuous variable to get quantized bit string and then correct error bits \cite{27}. Although sign reconciliation has the advantage of simplicity, it is only efficient in the case of high signal-to-noise ratio (SNR). Accordingly, Assche proposed a scheme called slice error correction (SEC) \cite{28,29}. Although SEC reconciliation can distill more than 1 bit per pulse in principle, but its secure transmission distance is limited to about 30 km. Leverrier subsequently proposed a promising scheme called multidimensional reconciliation \cite{30}, and it theoretically extended the secure transmission distance from 30 km to 50-100 km. Afterward, to pursue higher reconciliation speed or efficiency, multidimensional reconciliation scheme has been applied in several works with Polar code \cite{31}, LDPC code \cite{32,33,34,35}, and especially with Multi-edge type LDPC code (MET-LDPC) at low SNR regime \cite{18,19}. Whether Polar codes or LDPC codes are just applicable to some specific SNR. Since the imperfect of optical sources or other factors, the practical SNR might vary among transmissions significantly. When the practical SNR differs from the code's most suitable SNR, the reconciliation efficiency will be decreased. Therefore, some adaptive reconciliation algorithms with LDPC code are proposed in an attempt to emulate rateless operation and cope with the varying SNR by puncturing the LDPC codes \cite{18,34}. However, the punctured LDPC codes still perform high performance only in a small range. This forces the use of many different LDPC check matrices to cover a broad range of SNR in pursuit of high-performance reconciliation. As we all known, the matrix design of LDPC codes with high performance is also extremely difficult, especially for the long block length LDPC code which usually performs better than short code. In addition, a set of check matrices of long block length LDPC codes will forces Alice and Bob to consume lots of storage resources. With the fast increase in the repetition frequency of quantum channel, the storage resource of post-processing devices is facing unprecedented challenges, especially for the post-processing implementation based on FPGA \cite{36}.

In this paper, we propose a novel rateless reconciliation protocol for CV-QKD system based on Spinal code. Since the rateless operation, the proposed protocol can achieve a high-performance and adaptive reconciliation in a large range of SNRs and even the estimated SNR value differing from its true value of practical systems. In addition, we proved the security of the proposed protocol, which indicates that the privacy amplification is no more needed if the block size of reconciliation reaches $10^8$ bits. Being compared with the adaptive reconciliation schemes using the punctured LDPC codes, our proposed rateless reconciliation protocol does not need to design the check matrices for different SNRs, which greatly reduces the design cost and memory resource consumption. Simulation experiments show that the reconciliation efficiency of our protocol can achieve about $95\%$ for the ranging SNRs in a larger range $(0,0.5)$, even exceeds $96.5\%$ at extremely low SNR ($\le 0.03$) with almost $0\%$ FER. Therefore, our proposed protocol makes it much easier and stable for long-distance CV-QKD system to perform a high-performance and adaptive reconciliation.

The rest of this paper is organized as follows: In Sect. 2, the principle of spinal codes are introduced. In Sect. 3, we propose a novel reconciliation protocols for CV-QKD system based on spinal codes and discussed its security and performance. In Sect. 4, the simulation results and analysis of the novel reconciliation scheme are shown. Finally, conclusions are drawn in Sect. 5.

\section{Spinal Code}
The spinal code is a class of rateless codes \cite{37}, which has been proven to achieve Shannon capacity for the binary symmetric channel (BSC) and the additive white Gaussian noise (AWGN) channel with an efficient polynomial-time encoder and decoder \cite{38}. Compared with other coding schemes, such as LDPC code and Polar code, the spinal code is rateless code and has a simple coding structure, so it is more adaptable to variable channel. In order to make the proposed reconciliation algorithm easier to understand, we briefly introduce the encoding and decoding principle of spinal code in this section.
\subsection{Encoding spinal codes}
\label{sec:1}
The core of the spinal code is a hash function $g$, and a random number generator (RNG) known to both the transmitter and receiver. $g$ is chosen uniformly from a family of hash functions $\mathbb{G}$, which takes two inputs: a $v$-bits state and $k$ message bits, and returns a new $v$-bits state. That is,
\begin{equation}
g:{\{ 0,1\} ^v} \times {\{ 0,1\} ^k} \to {\{ 0,1\} ^v}
\end{equation}
The encoder maps $n$ bits input message $M = ({b_1},{b_2}, \cdots ,{b_n})$ to a stream of coded bits ${X_1},{X_2}, \cdots ,{X_{n/k}}$ using hash function and RNG. These coded bits are transmitted in sequence until the receiver signals that it is done decoding. The encoding process of Spinal codes is described as follows:
\begin{enumerate}[(1)]
\item Dividing $n$ bits input message $M$ into $n/k$ groups, each group of $k$ bits, and producing a sequence of $v$-bits states called the spine by applied $g$ sequentially to $m_i$. Let $m_i$ denotes bits ${b_{k(i - 1) + 1}},{b_{k(i - 1) + 2}}, \cdots ,{b_{ki}}$, so the sequence of states $s_i$ is simply generated as
\begin{equation}
{s_i} = g({s_{i - 1}},{m_i}),{\rm{  1}} \le i \le n/k,
\end{equation}

	where the initial state $s_0$ is known to both the encoder and decoder.
\item	Each of these $n/k$ spine values $s_i$  is used to RNG repeatedly and mapped to a longer binary sequence ${Q_i} = ({q_1}, \cdots ,{q_\tau }, \cdots )$. RNG is a function, which takes a $v$-bits state and another variable as inputs and generates a sequence of $\tau$ bits, as follow
    \begin{equation}
    RNG:{\{ 0,1\} ^v} \times \mathbb{N} \to {\{ 0,1\}^{\tau}}.
    \end{equation}
Because the requirements for RNG are similar to those for $h$, one suitable choice for RNG is to combine $h$ with a $v$-to-$\tau$-bits shift register \cite{37}.
\item Producing symbols $x_{i,\ell}$ as the $\ell $-th pass by acting a constellation mapping function $f$ on the $\ell$-th subsequence $({q_{(\ell  - 1)c + 1}}, \cdots ,{q_{\ell c}})$ of $Q_{i}$ with length of $c$ bits. The encoder continues generate additional symbols until the receiver decode the message $M$ successfully or both sides of communication to give up on the message.
\end{enumerate}

Let $b$ be the decimal number of a $c$-bits input to the constellation mapping. For the AWGN channel, the constellation mapping function $f$ encodes the $c$-bits input to a truncated Gaussian variable via the following mapping
\begin{equation}
f(b) = {\Phi ^{ - 1}}(\gamma  + (1 - 2\gamma )\alpha)\sqrt {P^*} ,
\end{equation}
where $\alpha = \frac{{2b + 1}}{{{2^{c{\rm{ + }}1}}}}$ ,and $\Phi (x)$ is the cumulative distribution function (CDF) of standard normal, $\gamma  = \Phi ( - \beta )$ limits the symbols in the range $\left[ { - \beta \sqrt {{P^*}} ,\beta \sqrt {{P^*}} } \right]$, and within that range the symbols are distributed like a Gaussian $N(0,P^*)$, $\beta$ controls the truncation width, $P^*$ is the modulation variance of truncated Gaussian variables.
 \begin{figure}[h]
\centering
  \includegraphics[width=12cm,height=8cm]{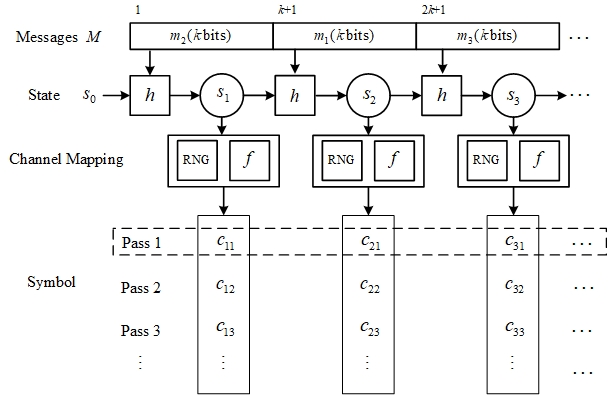}
\caption{Encoding process of Spinal code}
\end{figure}

\subsection{Decoding spinal codes - Bubble Decoding}
The central concept of decoding spinal codes is to search for the encoded message that differs least from the received signal over a tree of message prefixes. The decoding tree structure of the spinal code is shown as the figure. The root of the decoding tree is $s_0$, and corresponds to the zero-length message. Each node at depth $d$ corresponds to a prefix of length $kd$  bits, and is labeled with the final spine value  $s_d$ of that prefix. Every node has $2^k$ children, connected by edges  $e = ({s_d},{s_{d + 1}})$ representing a choice of $k$ message bits ${\bar m_e}$. As in the encoder, $s_{d+1}$ is $h({s_d},{\bar m_e})$. By walking back up the tree to the root and reading $k$ bits from each edge, we can find the message prefix for a given node. To the edge incident on node $s_d$, we assign a branch cost ${\left\| {{{\bar y}_d} - {{\bar x}_d}({s_d})} \right\|^2}$. Summing branch costs on the path from the root to a node gives the path cost of that node, equivalent to the sum in Eq.5.
\begin{equation}
\sum\limits_{i = 1}^{n/k} {{{\left\| {{{\bar y}_i} - {{\bar x}_i}({s_i})} \right\|}^2}}
\end{equation}

 \begin{figure}
\centering
  \includegraphics[width=12cm,height=7.3cm]{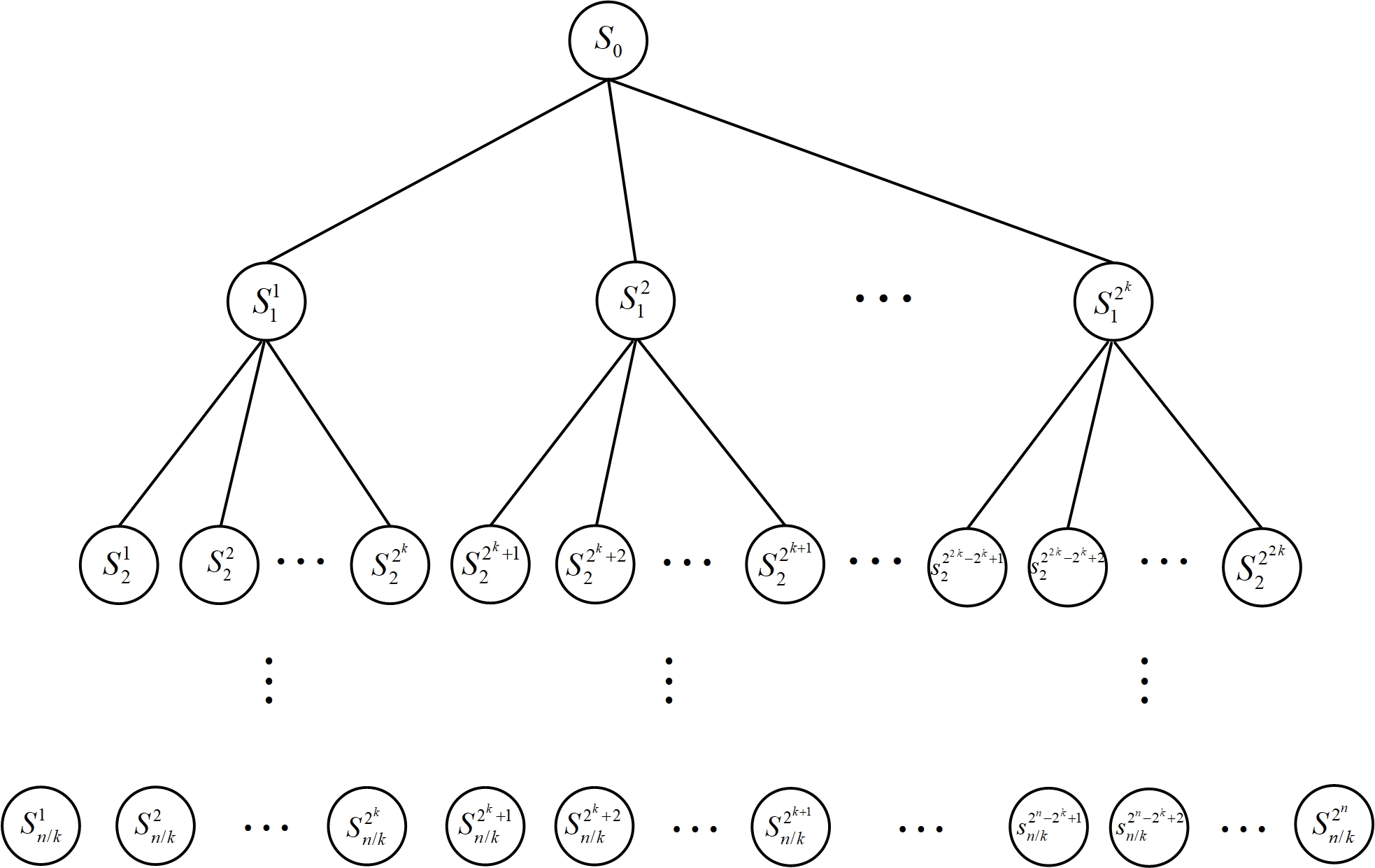}
\caption{Decoding tree of Spinal code}
\end{figure}

The Bubble algorithm is a strategy of pruning the decoding tree using greedy algorithm. The $B$ nodes with the smallest Euclidean distance are selected in each level of decoding tree, and the $B$ nodes form a set of candidate nodes. Extend the subtree rooted at these $B$ nodes. In decoding, each level needs to be expanded to a maximum of $B$ points. After calculation, $B \cdot {2^k}$ child nodes are generated, the candidate node set is cleared, $B \cdot {2^k}$ child nodes are sorted, and the optimal $B$ nodes are selected again. Put in the set of candidate nodes. The steps of the bubbling algorithm are as follows:

\begin{enumerate}[(1)]
\item Initialize the candidate node set and put the root node into the collection.
\item Expand all the nodes in the set once to get $B \cdot {2^k}$ nodes, and clear the candidate node set at this time. The data in the child node is updated, including the Euclidean distance, depth, state value, and path from the root node to the current node of the current node and the coded symbol receiving sequence.
\item Sort the $B \cdot {2^k}$ nodes according to the Euclidean distance, and put the B nodes with the smallest distance into the candidate node set.
\item If the depth of the point in the current candidate node set reaches $n/k$, the node with the lowest cost among the candidate nodes is selected as the final node of the decoding path, and the prefix in the final node corresponds to the decoding result, and the decoding ends. When the depth does not reach $n/k$, return to step 2.
\end{enumerate}

\section{Reconciliation Protocol based on Spinal Code}
In the previous section, we briefly introduced the spinal code. In this section, we propose a novel reconciliation algorithm for practical CV-QKD systems to distill common corrected keys from their correlated variables based on the spinal code. In order to analyze the security of the protocol, we prove its security from the perspective of information theory.
\subsection{Reconciliation Protocol}

The Gaussian-modulated coherent state CV-QKD is the most widely used scheme with the current techniques. We hence only discuss the reconciliation of CV-QKD with Gaussian modulation in this paper. After the quantum key establishment phase of a Gaussian-modulated coherent state CV-QKD system, Alice and Bob share weakly correlated continuous-variable raw data since the noises during the quantum transmission including physical noise and add noise from Eve. The noises can safely be assumed to be Gaussian since it corresponds to the case of the optimal attack for Eve \cite{39}. Let $X = ({x_1},{x_2}, \cdots )$ and $Y = ({y_1},{y_2}, \cdots )$ corresponding to correlated Gaussian vectors of Alice and Bob, respectively, then $Y=X+Z$ with $X \sim N{(0,{V_A})}$, $Z \sim N{(0,V_Z)}$ in the direct reconciliation case, and and $X = Y + Z'$ with $Y \sim N(0,{V_A}{\rm{ + }}{V_Z'})$, $Z' \sim N(0,{V_Z'})$ in the reverse reconciliation case, where $V_A$ is modulation variance and  $V_Z$, $V_Z'$ are noise variances. For long-distance CV-QKD protocols, the reverse reconciliation is required, which takes Bob's sequence as target keys to correct Alice's, i.e., Bob is the encoder and Alice is the decoder. Without loss of generality, we consider the reverse reconciliation.

Reviewing the traditional communication process with spinal code, it is easy to understand that the quantum transmission of the CV-QKD corresponds to the spinal symbols transmission of traditional communication. If we can transfer the noise of quantum transmission to the spinal symbols, then a virtual BWGN channel can be established for the reconciliation with spinal code. Subsequently, spinal code can be used for the reconciliation of CV-QKD.

However, the concrete noise values of correlated Gaussian vectors are not directly available. Fortunately, we design a way to securely and equivalently transfer the noise, i.e., Bob calculates the differences $\Delta =Y-C$ between his raw data $Y$  and spinal symbols $C$, and sends $\Delta$ to Alice over the classic authentication channel. Alice subsequently transfers the noise $Z$ of CV-QKD channel to spinal symbols $C$ by subtracting the received differences $\Delta$ from his raw data $X$ as the Eq.\ref{eq:6}, after which Alice obtains a noisy version $C'$ of $C$ as
\begin{equation}
\label{eq:6}
C' = X - \Delta  = C - Z.
\end{equation}
Since the symmetry of $Z \sim N(0, V_Z)$, $-Z$ follows the same distribution $-Z \sim N(0, V_Z)$.

By using the above method, we associate the correlated continuous-variable raw data with spinal codes. We now proceed to design a rate-compatible information reconciliation protocol for CV-QKD using these preparations as described above. At the beginning of reconciliation, the legitimate parties randomly choose and agree on a hash function. The schematic of the proposed protocol is plotted in Fig.3, of which the process is described in detail as follows.

$Step$ 1: Bob randomly generates a binary sequences $M$ of length $n$ as the secret key, and then encodes $M$ to spinal symbols $C = ({C_1},{C_2}, \cdots ,{C_{n/k}})$ through the spinal encoder.

In the encoding process, (i) spinal encoder firstly generates spine values $s_i =g(s_{i-1},m_i)$ of each subblock $m_i$; (ii) Then setting $s_i$ and ${t_j},j = 1,2, \cdots$ as seeds of RNG to obtain binary sequence $B_i$ and orderly acting function on subsequence of $B_i$ to generate spinal symbols ${C_i} = ({c_{i1}},{c_{i2}}, \cdots ,{c_{il}})$, where $s_0$ and $t_j$ are pre-shared secret data of Alice and Bob. The pass number $l$ is initially set as $l_{min}$ which calculated according to the practical SNR of the quantum channel as Eq.\ref{eq:7}. When the modulation variance of the Spinal encoder set as $P^*$, the SNR of virtual AWGN channel is ${S'_{NR}} = {P^{*}}/{V_Z}$. According to Shannon information theory, Bob needs to send at least $l_{min}$ passes for correcting successfully.

\begin{equation}
\label{eq:7}
{l_{\min }} = \left\lceil {\frac{k}{{0.5\log (1 + {{S'}_{NR}})}}} \right\rceil  = \left\lceil {\frac{k}{{0.5\log (1 + \frac{{{P^{*}}}}{{V_A}}{S_{NR}})}}} \right\rceil ,
\end{equation}
where $S_{NR}$ denotes the practical SNR of CV-QKD system, $\left\lceil x \right\rceil $ is the ceiling function, the log is to the base 2 and entropy is expressed in bits.

$Step$ 2: Bob calculates the differences $\Delta_i$ by subtracting his raw data ${y_1},{y_2},$ ${y_3}, \cdots $ with the spinal symbols ${c_{11}}, \cdots ,{c_{1l}},{c_{21}}, \cdots ,{c_{2l}}, \cdots $ orderly, and then sends the differences ${\Delta _i},i = 1,2, \cdots ,{{nl} \mathord{\left/
 {\vphantom {{nl} k}} \right.
 \kern-\nulldelimiterspace} k}$ to Alice over the classical authentic channel.

$Step$ 3: Alice subtracts the received differences  $\Delta_i$ from the corresponding raw data ${x_1},{x_2},{x_3}, \cdots $ on his hand to obtain the side information ${c'_{11}},{c'_{12}}, \cdots $, which are the noisy version of ${c_{11}},{c_{12}}, \cdots $. Then Alice uses the Bubble decoder to produces the estimated message $\hat M$ of $M$ with the prepared side information ${c'_{11}},{c'_{12}}, \cdots $ and the pre-shared $s_0$, $t_j, j=1,2,\cdots$.
\begin{equation}
\hat M = \mathop {\arg \min }\limits_{M' \in {{\{ 0,1\} }^n}} {\left\| {c' - \bar c(M')} \right\|^2},
\end{equation}
where $\bar c(X)$ is an encoder function that yields the vector of symbols for a message $X$.

$Step$ 4: Alice performs a cyclic redundancy check (CRC) check to verify the decoded message $\hat M$. If the CRC results of $\hat M$ and $M$ are equal, then the key string $M$ gets decoded successfully, otherwise Alice informs Bob to generates and sends additional difference information   until the latest updated message $\hat M$ pass CRC check or the decoding number reaches the predefined maximum number $i_{max}$.

The schematic diagram of the novel reconciliation protocol is shown as Fig.3. It is noted here that even if all CRC constraints are satisfied, there may exists undetected errors. But this situation rarely appears and can be neglected, since the probability of any other message $M'$ having likelihood higher than $M$ is made exponentially small in the message length $n$ [40], besides, collision of CRC also occurs with a extremely low probability.

 \begin{figure}
\centering
  \includegraphics[width=9.5cm,height=8cm]{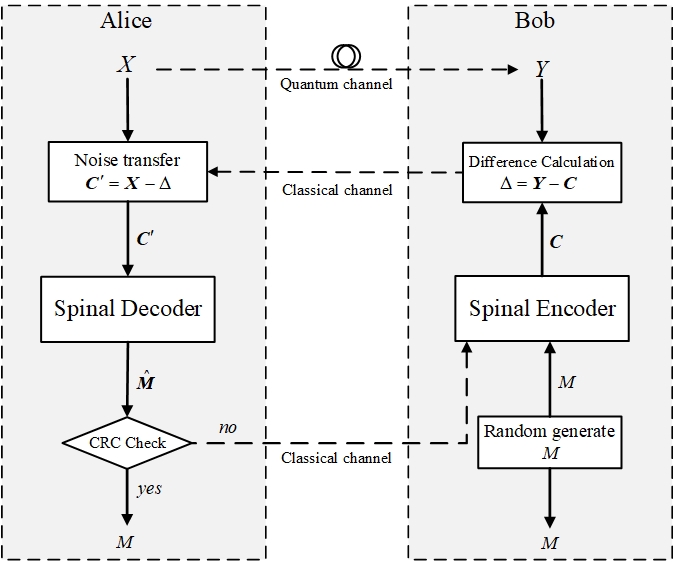}
\caption{Schematic diagram of the novel reconciliation scheme based on spinal code for CV-QKD}
\label{fig:1}
\end{figure}

\subsection{Performance analysis of protocol}
Now that the novel reconciliation protocol based on spinal code for CV-QKD has been proposed, it is necessary to consider the performance of our protocol. In this section, we investigate the performance of the proposed reconciliation protocol from two aspects including the security and reconciliation efficiency.

\subsubsection{Security analysis}

According to the proposed reconciliation protocol, the information that Eve obtains by monitoring the classic channel includes the differences $\Delta$ and check information $U$ used for CRC, which are together denoted as $\Gamma$. In addition, during the coherent states transmission over the quantum channel, Eve maybe obtain some information, which we note as $E$, about the correlated raw data. In order to discuss the security of the proposed reconciliation protocol, it is equivalent to investigate how much information about the key $M$ will leaked in the case that Eve knows $E$ and $\Gamma$. We have the conclusion presented in the following Theorem 1.

\textbf{Theorem 1} \emph{Let $M$ be a random $n$-bits string with uniform distribution, let $G$
be the hash function of spinal encoder chosen at random from a family of hash functions $\mathbb{G\rm{:}}{\{ 0,1\} ^v} \times {\{ 0,1\} ^k} \to {\{ 0,1\} ^v}$, $S$ and $T$ be the random initial state of the hash function $G$ and one seed of RNG in the spinal encoder respectively, and let $E$ and $T$ denote the information Eve obtained through the quantum channel and classic channel respectively, and $U$ denotes the $\lambda$ bits check codes used for CRC. If Alice and Bob share secret random strings $s = {{\rm{\{ 0,1\} }}^v}$  and $t = {{\rm{\{ 0,1\} }}^\omega }$ as the starting states of hash function and RNG of the proposed reconciliation respectively, then Eve's expected information about $M$, given $\Gamma$ and $E$, satisfies \[I(M;G,\Gamma ,E) < \frac{n}{{{2^{v + \omega }}}}+\lambda\]}

\emph{Proof} According to the chain rule of mutual information, we expand the leaked information $I(M;G,\Gamma ,E)$ as
\begin{equation}
I(M;G,\Gamma ,E) = I(M,S,T;G,\Gamma ,E) - I(S,T;G,\Gamma ,E|M)
\end{equation}
Since the mutual information is non-negative, we have
\begin{equation}
I(M;G,\Gamma ,E) \le I(M,S,T;G,\Gamma ,E)
\label{eq:10}
\end{equation}
We now discuss $I(M,S,T;G,\Gamma ,E)$ to indirectly derive the upper bound of $I(M;G,\Gamma ,E)$. Similarly, we use chain rule of mutual information, it can be express as
\begin{equation}
I(M,S,T;G,\Gamma ,E) = I(S,T;G,\Gamma ,E) + I(M;G,\Gamma ,E|S,T)
\end{equation}
Since the spinal codes sequentially apply hash function and RNG to the message bits to produce coded symbols for transmission, it presents a framework for making Shannon's random coding ideas, i.e., the output of spinal encoder is random and sensitive to input. It ensures that two input of spinal encoder that differ in even one bit lead to very different coded symbols \cite{37}. Therefore, even if $G,\Gamma ,E$ is given, Eve still knows nothing about $S$ and $T$ without knowing $M$. In addition, $S$ and $T$ are random strings. Then, we have $I(S,T;G,\Gamma ,E) = 0$. Averaging over values of $S$ and $T$, we can further obtain that
\begin{equation}
\begin{split}
I(M,S,T;G,\Gamma ,E) &= I(M;G,\Gamma ,E|S,T)\\
 &= P(S = s,T = t)I(M;G,\Gamma ,E|S = s,T = t)\ + \\
 &\sum\limits_{\begin{array}{*{20}{c}}
{x \in {{{\rm{\{ 0,1\} }}}^v}\backslash s}\\
{y \in {{\{ 0,1\} }^\omega }\backslash t}
\end{array}} {P(S = x,T = y)I(M;G,\Gamma ,E|S = x,T = y)} ,
\end{split}
\label{eq:12}
\end{equation}
where ${{{\rm{\{ 0,1\} }}}^v}\backslash s $ denotes the set excluding $S=s$. $P(S,T)$ is the probability distribution of $S$ and $T$. Because $S$, $T$ are random and independent strings, for any $x \in {\left\{ {0,1} \right\}^v}$, $y \in {\{ 0,1\} ^\omega }$, $x$ and $y$ are equally likely candidates for $S$ and $T$. Then, $P(S = x,T = y) = {1 \mathord{\left/
 {\vphantom {1 {{2^{v + \omega }}}}} \right.
 \kern-\nulldelimiterspace} {{2^{v + \omega }}}}$.

We first discuss the first term on the left side of Eq.\ref{eq:12}. From the perspective of Eve, if he knows $S=s$ and $T=t$, it is equivalent to establishing a virtual channel with signal-to-noise ratio $S_{NR}^e$, where the differences $ - \Delta  = C - Y$ can be regarded as a noise version of the spinal symbol $C$ added noise $Y$. Therefore, using Shannon information theory and conditional mutual information formula, $I(M;G,\Gamma ,E|S = s,T = t)$ can be expanded and bounded as
\begin{equation}
\begin{split}
I(M;G,\Gamma ,E|S = s,T = t) &= H(M|S = s,T = t) - H(M|S = s,T = t,G,\Gamma ,E)\\
 &\le H(M) - H(M|S = s,T = t,G,\Gamma ,E)\\
 &= n - \max \{ n - \frac{N}{2}\log (1 + S_{NR}^e),0\} \\
 &\le n,
\end{split}
\end{equation}
where $N$ is the length of differences $\Delta$. $S_{NR}^e = \frac{{{P^*}}}{{{V_B}}}$, where $P^*$ and $V_B$ denote the modulation variance of the spinal symbol $C$ and Bob's measured raw data $Y$, respectively.

Next, we discuss the second item. For $u \in {\left\{ {0,1} \right\}^{\lambda}}$, let $c_u$  be the number of $m \in {\left\{ {0,1} \right\}^n}$  that are consistent with $u$, i.e., satisfying $CRC(m) = u$. Since $M$ is random string, then $P(U=u)= c_u/2^n$. As mentioned in the previous, the output of spinal codes is random and sensitive to input \cite{37}. For two different inputs, their outputs of the spinal codes are independent of each other \cite{38}. Hence, Eve cannot learn any information about $M$ using his obtained information $G,\Gamma ,E$ when $x \ne s$ and $y \ne t$, i.e., for any $x \ne s$ and $y \ne t$, all consistent $m$ with $CRC(m) = u$ are equally likely candidates, then $P(M|S = x,T = y,U = u,G,E,\Delta ) = \frac{1}{{{c_u}}}$. Therefore, we have
\[I(M;G,\Gamma ,E|S = x,T = y) = H(M) - H(M|S = x,T = y,G,\Gamma ,E)\]
\begin{equation}
\begin{split}
 &= n - \sum\limits_{u \in {{\{ 0,1\} }^\lambda }} {P(U = u)} H(M|S = x,T = y,U = u,G,E,\Delta )\\
 &= n - \sum\limits_{u \in {{\{ 0,1\} }^\lambda }} {\frac{{{c_u}}}{{{2^n}}}} \log {c_u}\\
 &= \sum\limits_{u \in {{\{ 0,1\} }^\lambda }} {\frac{{{c_u}}}{{{2^n}}}} \log \frac{{{2^n}}}{{{c_u}}}\\
 &\le \lambda.
 \end{split}
\end{equation}
According to the above discussion, then we derived that
\begin{equation}
\begin{split}
I(M,S,T;G,\Gamma ,E) &\le \frac{n}{{{2^{v + \omega }}}} + \sum\limits_{x \in {{\{ 0,1\} }^v}\backslash s,y \in {{\{ 0,1\} }^\omega }\backslash t} {\frac{\lambda }{{{2^{v + \omega }}}}} \\
& < \frac{n}{{{2^{v + \omega }}}} + \lambda ,
\end{split}
\label{eq:15}
\end{equation}
combining with Eq.\ref{eq:10}$\sim$Eq.\ref{eq:15}, we finally get
\begin{equation}
I(M;G,\Gamma ,E) < \frac{n}{{{2^{v + \omega }}}}+\lambda.
\end{equation}

The Theorem1 indicates that, if we distill $ n $-bits corrected keys $ M $ using the proposed reconciliation protocol, Eve can obtain $\frac{n}{{{2^{v + \omega }}}}{\rm{ + }}\lambda $  bits partial information about $ M $ at most. One can eliminate the leaked information in privacy amplification stage by performing a universal hash function on a larger body of corrected keys at a compression rate $r \approx 1 - (\frac{\lambda }{n} + \frac{1}{{{2^{v + \omega }}}})$. After that, the partially secret keys can be distilled to the highly secret keys which can be used directly as finial secret keys.

\subsubsection{Reconciliation efficiency analysis}
Let us now discuss the proposed protocol's reconciliation efficiency, which is a significant indicator to evaluate the performance of the information reconciliation step. Because our works focus on CV-QKD system with Gaussian modulation. We have to take into account that our quantum channel is Gaussian, and then the reconciliation efficiencies with respect to Gaussian channel capacity is computed as follow
\begin{equation}
\beta  = \frac{R}{C(S_{NR})},
\end{equation}
where $R$ is the rate of reconciliation protocol and $C(S_{NR}) = \frac{1}{2}\log (1 + {S_{NR}})$ is the classical capacity of the quantum channel for Gaussian variables.

The code rate of the spinal codes in traditional communication is $R' = \frac{n}{{(Ln/k)}}$, where $L$ is the passes number of spinal symbols transmission. $L$ is not fixed, since spinal codes are able to take advantage of channel variations and the produces the spinal symbols to receiver at a higher code rate prior to a lower code rate. This is why spinal code is rateless code. However, the traditional communication does not need to consider the leakage and consumption of the message $M$. Therefore, we should define the $R$ with considering the security. Next, let us investigate the code rate $R$ of our proposed reconciliation protocol.

 In our proposed protocol, the maximum leaked information about the correct keys $M$ is $n/2^{v+\omega}+\lambda$  according to the Theorem 1. This is based on the pre-shared secure data $s$ and $t$ of total length  $v+\omega$ bits, which needs to consume the final secret keys. We know that the final secret key is converted from the correct key at a certain compress-ratio $r$ by performing a privacy amplification, i.e., it is indirectly equivalent to consume  $(v+\omega)/r$ bits correct keys for reconciliation. Hence, the remaining correct keys of the proposed protocol are $n-(n/2^{v+\omega}+\lambda)-(v+\omega)/r$ bits with considering the security, and these correct keys are extracted from $nL/k$ bits correlated raw data of the legitimate parties. Therefore, the code rate $R$ of the proposed reconciliation protocol is given by 
\begin{equation}
R = \frac{{k\left[ {n - ({n \mathord{\left/
 {\vphantom {n {{2^{v + \omega }}}}} \right.
 \kern-\nulldelimiterspace} {{2^{v + \omega }}}} + \lambda ) - {{(v + \omega )} \mathord{\left/
 {\vphantom {{(v + \omega )} r}} \right.
 \kern-\nulldelimiterspace} r}} \right]}}{{nL}}.
\end{equation}

In practical systems, for a fixed optical transmission distance between Alice and Bob, the reconciliation efficiency can be optimized by tuning the modulation variance of spinal codes. Now, let us investigate the adjustable range of modulation variance $P^*$. Obviously, the code rate $R$ reaches its maximum when $L=l_{min}$, and then the reconciliation efficiency $\beta$ achieves maximum in this case. According to the Shannon information theory, it is satisfied that
\begin{equation}
\frac{{2k\left[ {n - ({n \mathord{\left/
 {\vphantom {n {{2^{v + \omega }}}}} \right.
 \kern-\nulldelimiterspace} {{2^{v + \omega }}}} + \lambda ) - {{(v + \omega )} \mathord{\left/
 {\vphantom {{(v + \omega )} r}} \right.
 \kern-\nulldelimiterspace} r}} \right]}}{{n{l_{\min }}\log (1 + {S_{NR}})}} \le 1,
 \label{eq:19}
\end{equation}
as we discussed previous, $n/2^{v+\omega}$ is very small and then can be ignored, and $\omega  = \frac{{w{l_{\min }}c}}{v}$, where $w$ is the length of another input $\mathbb {N}$ of RNG. Then, the Eq.\ref{eq:19} is approximately expressed as
\begin{equation}
n - \lambda  - \frac{{{v^2}\log (1 + {S_{NR}}^\prime ) + 2kwc}}{{vr\log (1 + {S_{NR}}^\prime )}} \le n\frac{{\log (1 + {S_{NR}})}}{{\log (1 + {S_{NR}}^\prime )}},
\label{eq:20}
\end{equation}
where ${S_{NR}}^\prime  = \frac{{{P^*}}}{{{V_Z}}}$. By solving Eq.\ref{eq:20}, we can get the range of the modulation variance of spinal symbols as follow
\begin{equation}
{P^*} \le \frac{{({2^\eta } - 1){V_A}}}{{{S_{NR}}}},
\end{equation}
where $\eta  = \frac{{vrn\log (1 + {S_{NR}}) + 2kcw}}{{vr(n - v/r - d)}}$.

\section{Simulation and analysis}
In this section, the simulation experiments are performed to show the performance of the proposed reconciliation protocol. Also, the comparative simulations between reconciliation scheme in \cite{18} and our scheme are carried out.

\subsection{Parameter setting}

\textbf{Hash function $h$}. Spinal codes rely on the mixing ability of the hash function to provide the output independence. The existing research works in \cite{18} simulated the performance of three widely used hash function Salsa20, lookup3.1 and one-at-a-time, and showed no discernible difference in performance between these three hash functions in spinal codes. But the processing speed of one-at-a-time hash function is fastest compared with the other hash functions. We used one-at-a-time in our experiments.

\textbf{RNG}. Since the hash function possesses the mixing ability, we use one-at-a-time to implement RNG. The encoder and decoder call $g(s_i, t_j)$ to generate longer output symbols. This method has the desirable property that not every output symbol has to be generated in sequence, i.e., if some frames containing symbols are not recovered, the decoder need not generate the missing symbols. Block length $n$. The longer block length $n$ lead a higher reconciliation efficiency. However, the complexity of reconciliation scheme will increase with the block length. In our experiments, we set the block length $n= 1024$.

\textbf{Picking $k$, $B$ and $c$}. We know that $k$ and $B$ have the greatest impact on determining the complexity of the reconciliation. Larger values of $k$ don't do well at low compute budgets, but smaller values of $k$ underperform at higher SNRs. Each decoder can use a value of $B$ according to its price performance ratio. We set $k=4$, $B=256$, $c=6$ in our experiments referring to the literature \cite{18}.

\textbf{SNR}. A particular challenge of long distance CV-QKD systems is achieving the high-performance reconciliation at low SNR (normally lower than $0.5$) of the quantum channel. So we only measure performance across an SNR range $(0,0.5)$ in our experiments.

\textbf{Other parameters}. In fact, for the fixed noise variance, changing modulation variance $V_A$ leads to different SNR values, i.e., it is equivalent to adjusting the $V_A$ by taking different value of SNR. Without loss of generally, we set modulation variance of Alice as $V_A=1$. The modulation variance of truncated Gaussian variable (spinal symbol) is set as ${P^*}{\rm{ = }}\frac{{({2^\eta } - 1){V_A}}}{{{S_{NR}}}}$, the maximum iteration number is set as $i_{max}=50$.

\subsection{Experiments results and analysis}
We mainly display the reconciliation performance of the proposed information protocol in terms of reconciliation efficiency, frame error rate (FER) and iteration number of decoding, which are important indicators to evaluate the reconciliation performance. Since reference \cite{18} only listed the experimental results of the reconciliation efficiency, the comparative experiments between scheme in \cite{18} and our scheme are only carried out for the reconciliation efficiency. To the best of our knowledge, performance of the existing reconciliation scheme \cite{13,17,18} is normally simulated at a given SNR, i.e., assuming the SNR of quantum channel is accurately known to the communication parties. However, as we all known, the SNR of quantum channel in practical CV-QKD systems is estimated by the parameter estimation stage. Due to the finite-size effect, there exist statistical fluctuation in parameter estimation, and then it is difficult to accurately estimate the true value of SNR. Take this situation into account, we performed some additional experimental simulations at the estimated SNR value differs from its true value. All the experiments at each SNR average the performance over 100 data blocks.

\subsubsection{Reconciliation efficiency}
\begin{figure}[h]
\centering
  \includegraphics[width=12cm, height=6.5cm]{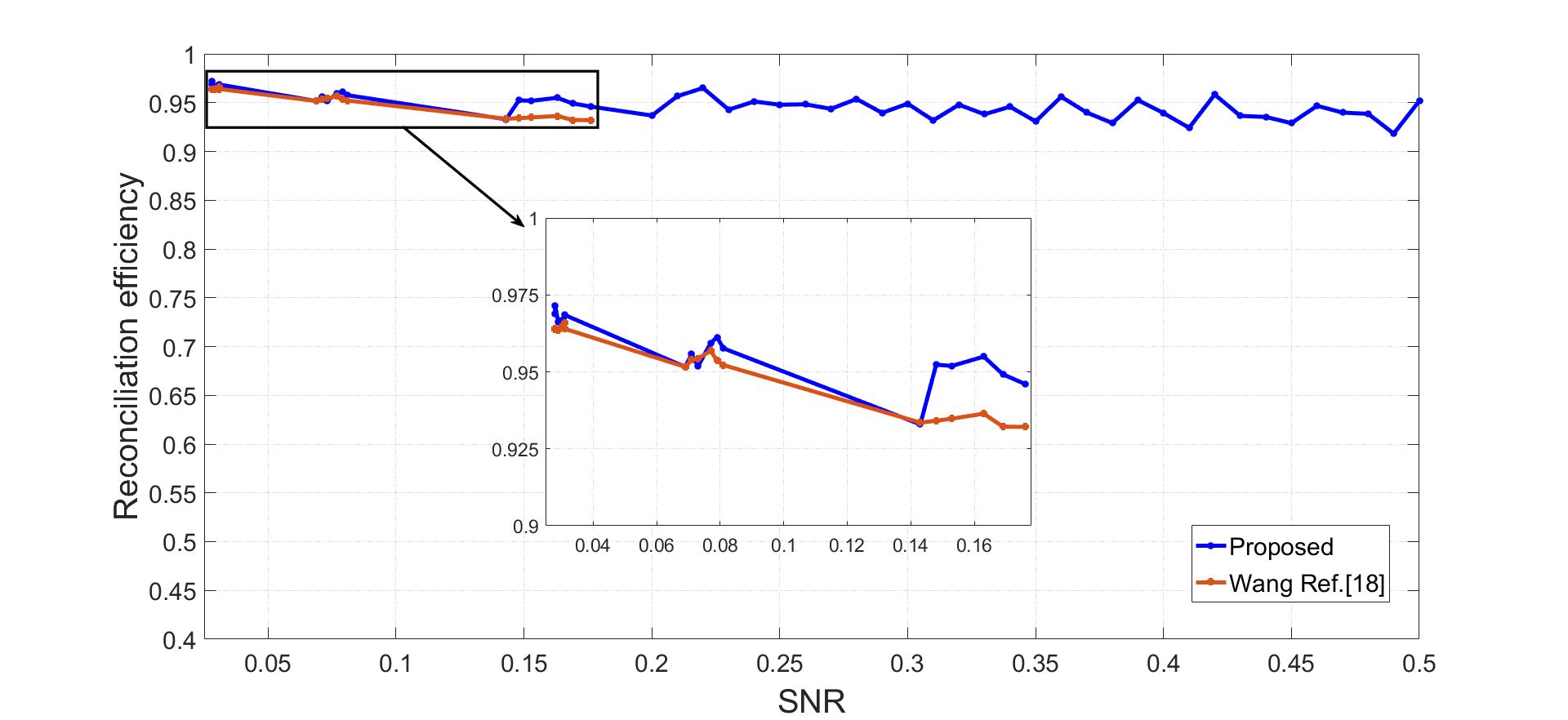}
\caption{Reconciliation efficiency of the proposed reconciliation protocol}
\label{fig:3}
\end{figure}

Table.\ref{tab:1} gives a comparison of the reconciliation efficiency between the scheme in reference \cite{18}  and our proposed scheme. According to the table, it can be seen that our scheme generally performs better than the reference \cite{18} almost for every SNR. In addition, the reconciliation efficiencies of our scheme in a larger range of SNRs is given by Fig.\ref{fig:3}, which shows that our scheme always achieve the reconciliation efficiency stably around $95\%$.

These simulation results indicate that our solution can achieve a high-performance and stable adaptive reconciliation for varying SNR. This is because our scheme is a rateless reconciliation protocl and always produce the symbols at a higher code rate prior to a lower code rate.
\begin{table}
\centering
\caption{Reconciliation efficiency comparison between scheme in Ref.\cite{18}  and proposed scheme}
\label{tab:1}
\begin{tabular}{p{1.5cm}p{1.5cm}p{1.5cm}|p{1.5cm}p{1.5cm}p{1.5cm}}
\hline\noalign{\smallskip}

SNR&Proposed&Ref.[18]&SNR&Proposed&Ref.[18]\\
\noalign{\smallskip}\hline\noalign{\smallskip}
0.0277  & 96.89\% & 96.40\% & 0.077 & 95.94\% & 95.68\%\\
0.0280  & 97.16\% & 96.38\% & 0.079 & 96.13\% & 95.36\%\\
0.0286  & 96.62\% & 96.36\% & 0.081 & 95.76\% & 95.22\%\\
0.0299  & 96.68\% & 96.46\% & 0.143 & 93.29\% & 93.35\%\\
0.0306  & 96.59\% & 96.59\% & 0.148 & 95.23\% & 93.41\%\\
0.0314  & 96.85\% & 96.40\% & 0.153 & 95.19\% & 93.48\%\\
0.0690  & 95.16\% & 95.16\% & 0.163 & 95.50\% & 93.64\%\\
0.0710  & 95.58\% & 95.39\% & 0.169 & 94.92\% & 93.22\%\\
0.0730  & 95.18\% & 95.43\% & 0.176 & 94.60\% & 93.21\%\\
\noalign{\smallskip}\hline
\end{tabular}
\end{table}

 \begin{figure}[h]
\centering
  \includegraphics[width=12cm,height=6cm]{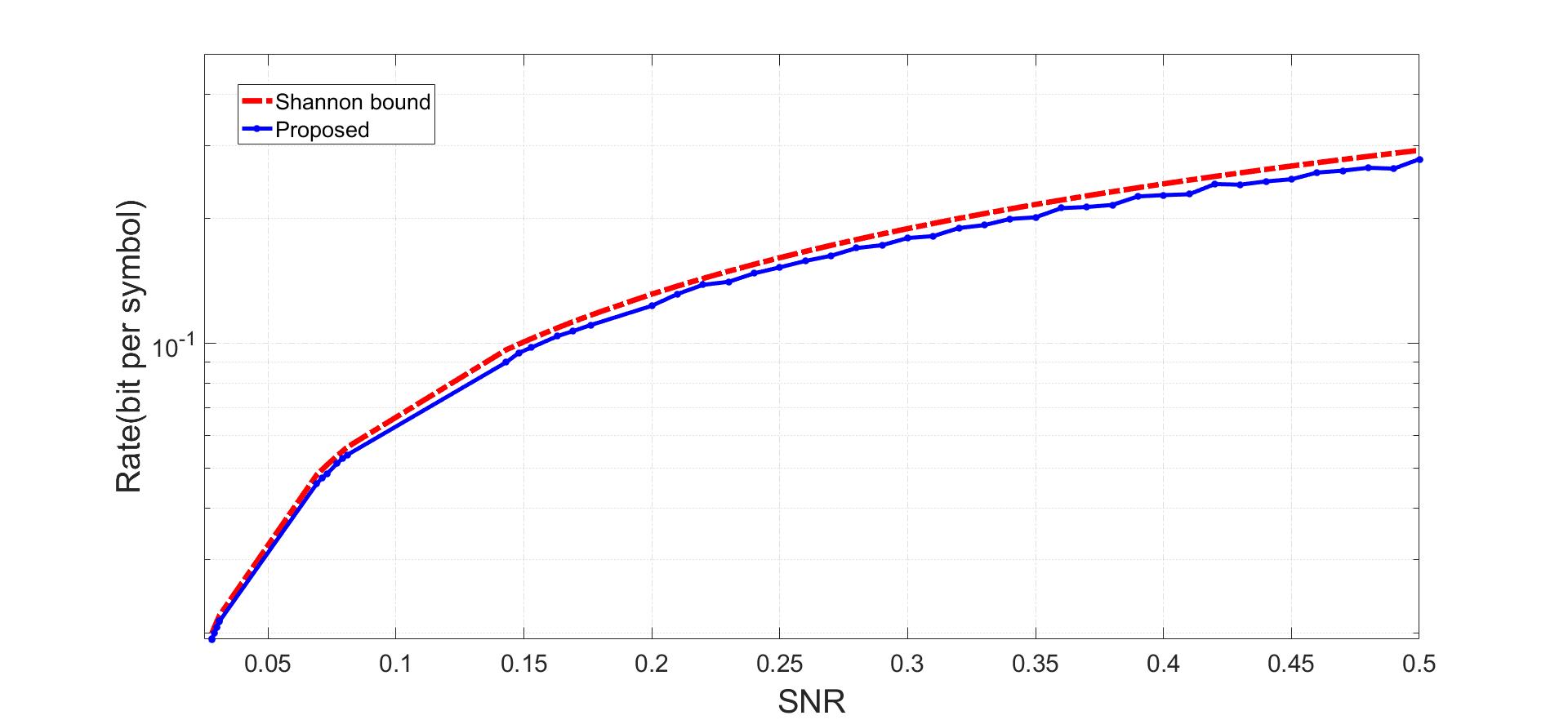}
\caption{The proposed reconciliation protocol approaches to Shannon bound}
\end{figure}

\subsubsection{Frame error rate}
\begin{figure}[h]
\centering
  \includegraphics[width=12cm,height=6cm]{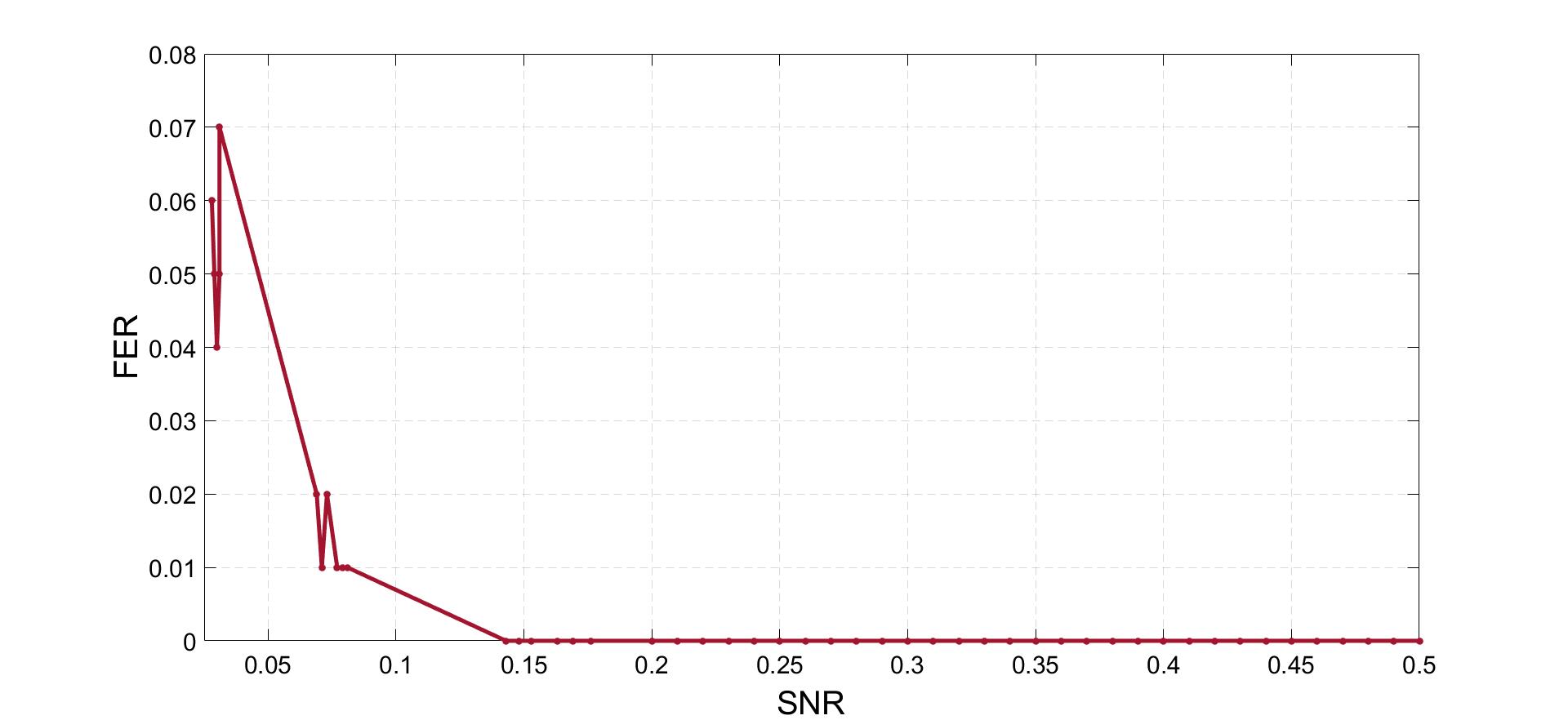}
\caption{Frame error rate of the proposed reconciliation protocol}
\label{fig:6}
\end{figure}

The FER refers to the failure probability of reconciliation, which has a great influence on the secret key rate of CV-QKD system. It is also an important indicator of reconciliation performance. Fig.\ref{fig:6} shows the FER of the proposed reconciliation protocol. It indicates that our reconciliation protocol can finish the adaptive reconciliation with almost $100\%$ success rate.
\subsubsection{Iteration number of decoding}
\begin{figure}[h]
\centering
  \includegraphics[width=12cm,height=6cm]{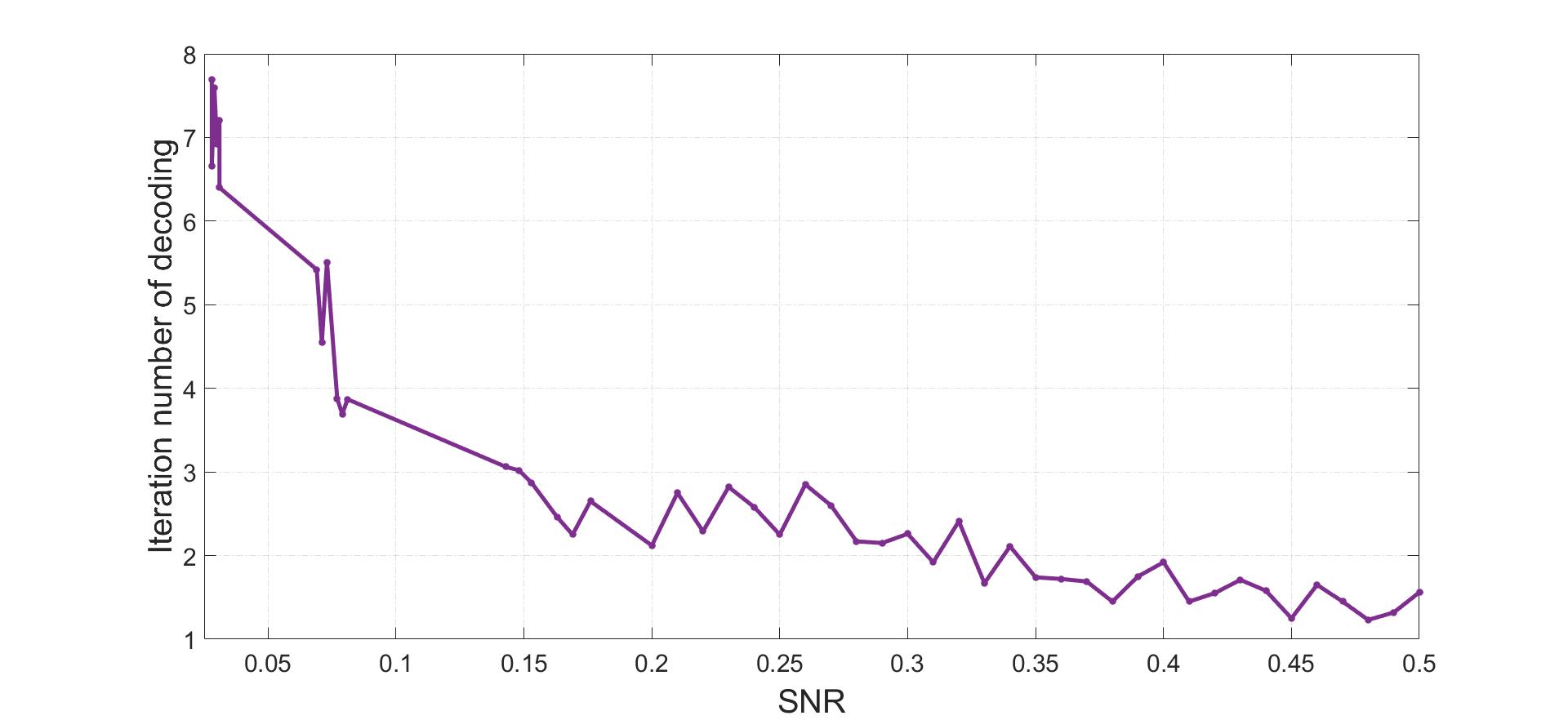}
\caption{Iteration number of decoding of the proposed reconciliation protocol}
\label{fig:5}
\end{figure}

We simulate and record the average iteration number of a data block decoding to reflect the decoding speed of our reconciliation algorithm indirectly. Fig.\ref{fig:5} shows that the proposed reconciliation scheme finish the reconciliation only requires a very small number of iterations.

It is noted that, although the total decoding cost scales of one iteration is ${\rm O}(nB{2^k}(k + \log B + v))$, the cost calculation of the $B\cdot2^k$ candidate nodes of each depth is independent of each other. Therefore, decoder of spinal codes is very suitable for using parallel algorithms to achieve competitive throughput. But accelerating the throughput of decoder is currently not considered in this paper.

\section{Conclusions}
In this paper, we propose a novel rate-adaptive information reconciliation protocol for practical CV-QKD systems based on spinal code. Compared with the existing adaptive reconciliation algorithm based on the LDPC code, our proposed protocol does not need to design the check matrices for different SNRs, which greatly reduces the design cost and memory resource consumption. In addition, we proved the security of the proposed protocol based on information theory. Simulation experiments show that the reconciliation efficiency of our protocol can achieve about $95\%$ for the ranging SNRs. That is to say, our protocol can realize adaptive and high-efficiency information reconciliation under the premise of ensuring security.


\bibliographystyle{unsrt}
\bibliography{template}

\end{document}